\documentclass[a4paper,11pt]{article}
\pdfoutput=1 

\usepackage{jinstpub} 

\usepackage{lineno}

\usepackage[numbers,sort&compress]{natbib} 

\usepackage{xcolor}
\definecolor{light-gray}{gray}{0.95}
\definecolor{dark-green}{rgb}{0,0.5,0}
\definecolor{lightblue}{rgb}{.80,.90,1}
\definecolor{lightred}{rgb}{1, 0.8, 0.8}
\newcommand{\code}[1]{\colorbox{light-gray}{\texttt{#1}}}

\usepackage{soulutf8}

\usepackage[section]{placeins}

\newcommand{\VS}[1]{\textcolor{blue}{#1}}
\newcommand{\AT}[1]{\textcolor{dark-green}{#1}}
\newcommand{\AW}[1]{\textcolor{orange}{#1}}

\title{\boldmath GEANT4 simulation and spectrum restoration of a pixelated X-ray detector}

\author[a]{Andrii Tykhonov}
\author[b,c]{Alexander Winkler}
\author[d]{Volodymyr Smoliar}

\affiliation[a]{Department of Nuclear and Particle Physics, University of Geneva, CH-1211, Switzerland}
\affiliation[b]{Helsinki Institute of Physics, Gustaf H\"{a}llstr\"{o}min katu 2, P.O. Box 64, FI-00014 University of Helsinki, Finland}
\affiliation[c]{Detection Technology Plc, Ahventie 4 B, FI-02170 Espoo, Finland}
\affiliation[d]{Department of Theoretical and Experimental Nuclear Physics, Odessa Polytechnic National University, Odessa, Ukraine}

\emailAdd{andrii.tykhonov@unige.ch}

\emailAdd{alexander.winkler@helsinki.fi} 

\emailAdd{svp@op.edu.ua}

\newcommand{\micron}{$\text{\textmu m}$}

\abstract{We perform a detailed simulation of a pixelated CdTe detector using the GEANT4 toolkit completed with a custom code emulating the detector's electronic response. We demonstrate that a measured tungsten X-ray spectrum can be majorly restored back to the original incident spectrum using the developed model, without requiring the dedicated hardware charge sharing correction.}

\makeatletter
\gdef\@fpheader{}
\makeatother

\begin{document}
\maketitle
\flushbottom

\section{Introduction}

In this work, we develop simulation software for X-ray detectors. We also suggest an alternative solution for the commonly used charge sharing correction algorithm C8P1~\citep{Maj_2012} that is required for photon counting detectors with several energy bins and small pixels, to maintain good energy resolution~\citep{Maj_2012, Ullberg_2013}. The simulation software has two main components: (i) simulation of physics processes, including X-ray propagation and absorption, as well as secondary emission and fluorescence in the sensitive material and other materials of the detector setup; (ii) simulation of the detector response, including the conversion of the absorbed energy into electron-hole pairs, charge sharing, ASIC non-linearity, electronic noise, and energy resolution. The first component is implemented in the framework of GEANT4 toolkit version \mbox{\code{4.11.0.0}}~\citep{geant4,Allison:2006ve,Allison:2016lfl}. The second one is implemented as a custom in-house code, written in Python\footnote{https://docs.python.org/3/library/}. In addition to the simulation software, a new approach to replace the charge sharing correction in the X-ray detectors is studied and optimized based on the developed simulation software.

This work is structured as follows. In Section~\ref{sec:geant4}, the developed X-ray simulation setup in the framework of the GEANT4 toolkit is briefly described. The simulation of the detector response and comparisons with real photon counting detector data 
are presented in Section~\ref{sec:response}. 
The spectrum restoration procedure is described and tested with both monochromatic and X-ray tube spectra in Section~\ref{sec:unfolding}. Discussion and conclusions are then provided in Section~\ref{sec:conclusions}.

\section{Simulation of X-ray propagation}
\label{sec:geant4}

\begin{figure}[b!]
\begin{center}
\includegraphics[width=0.4\textwidth]{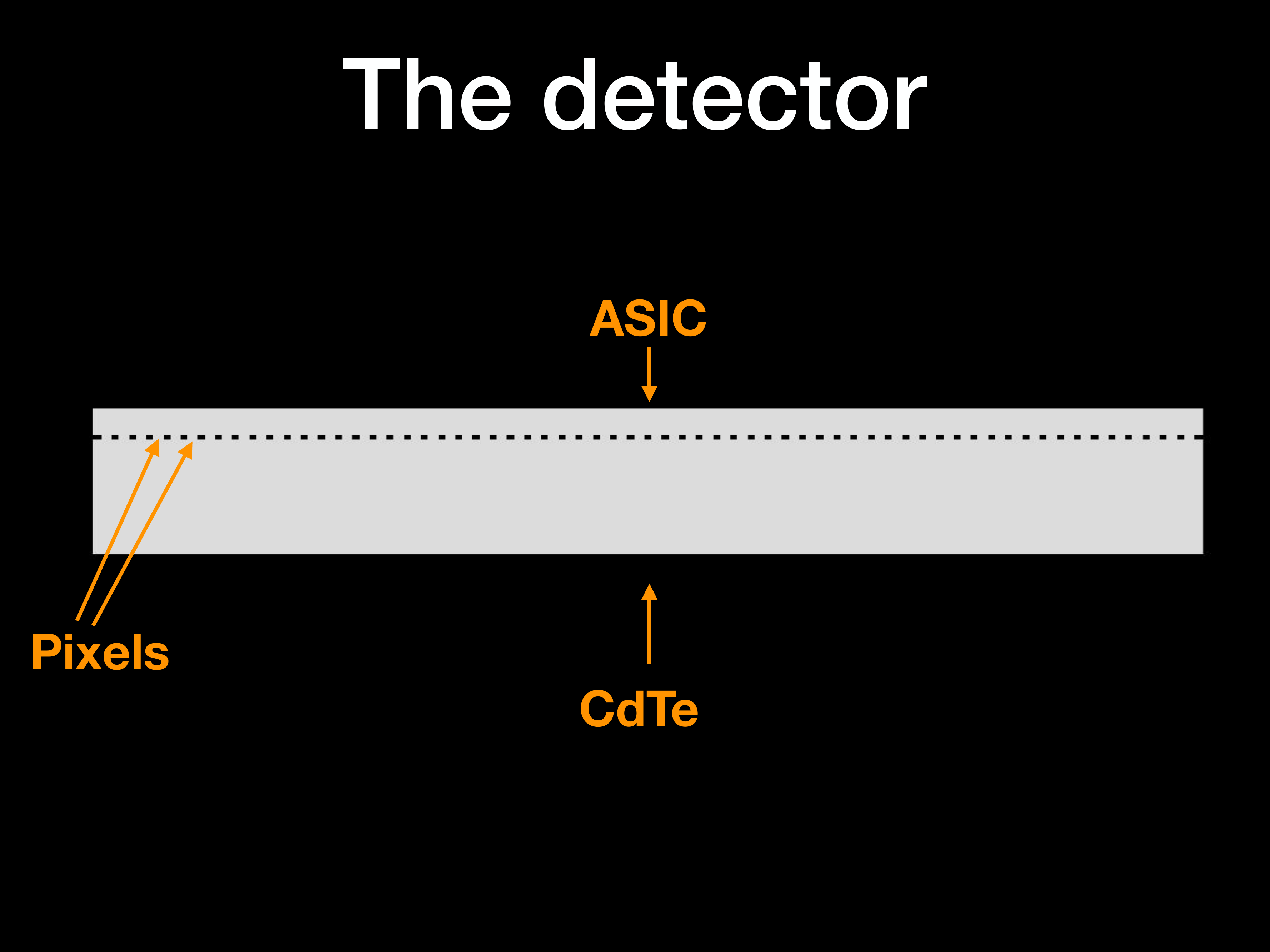}
\includegraphics[width=0.4\textwidth]{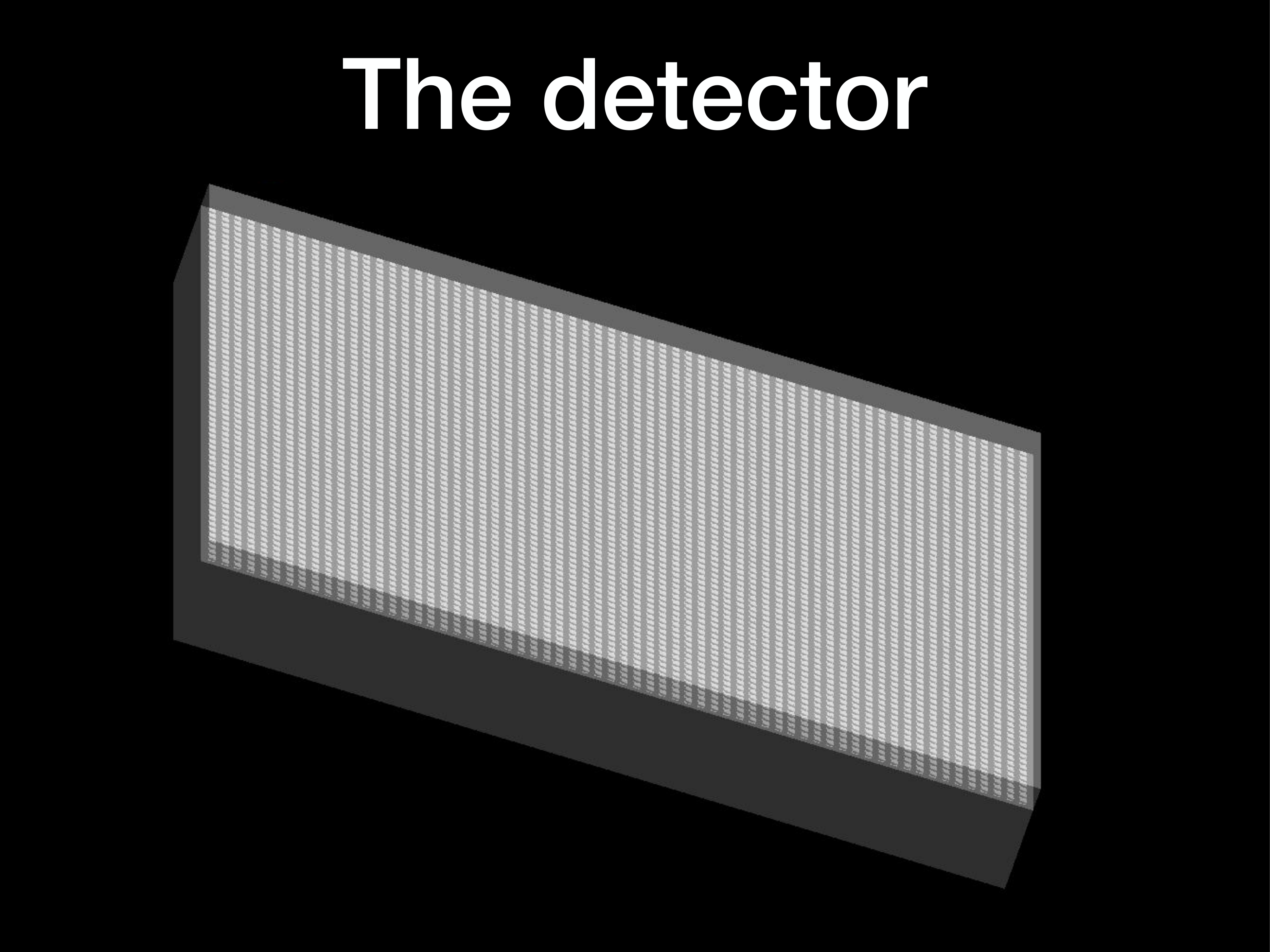} \\
\includegraphics[width=0.4\textwidth]{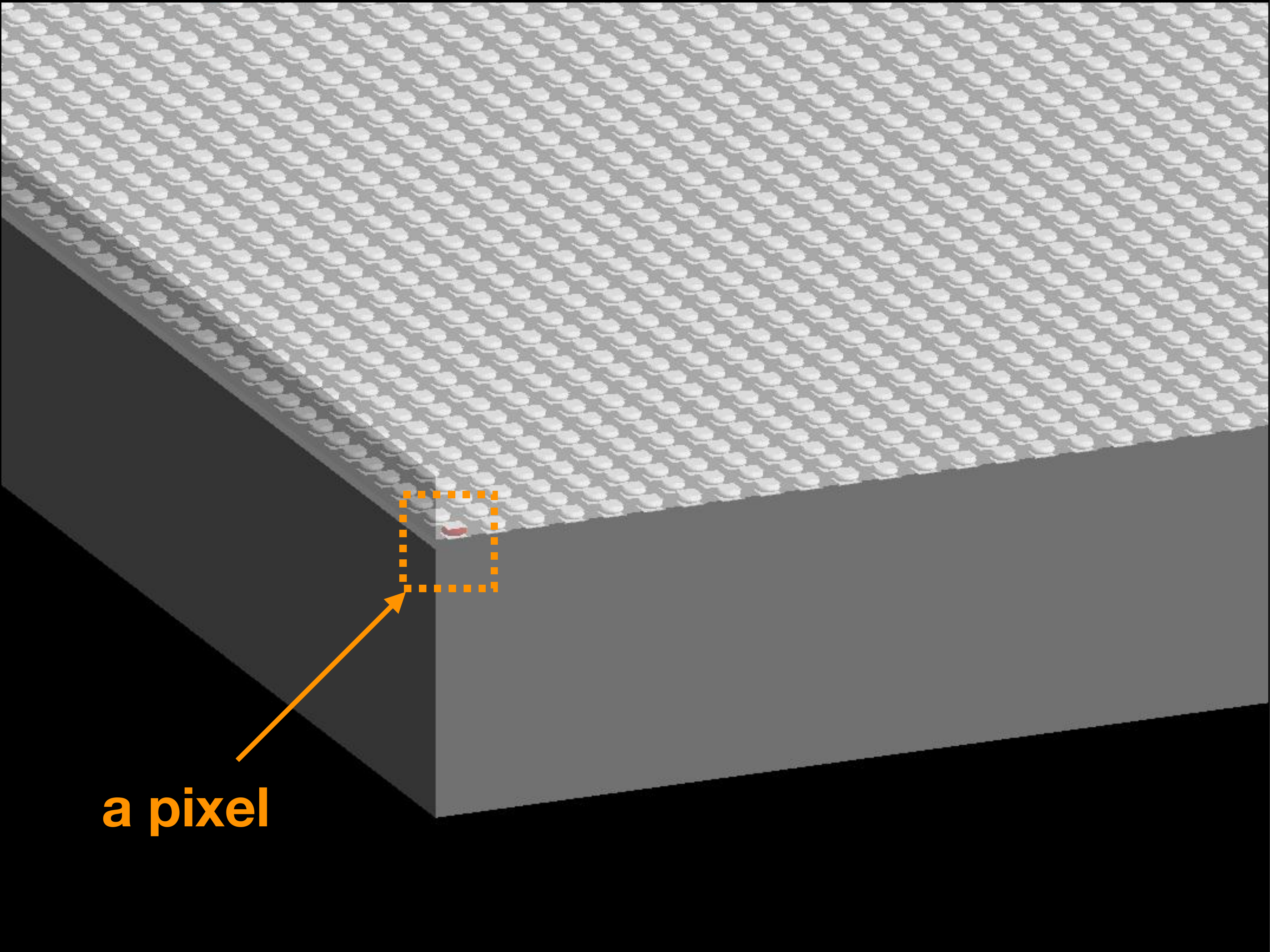}
\includegraphics[width=0.4\textwidth]{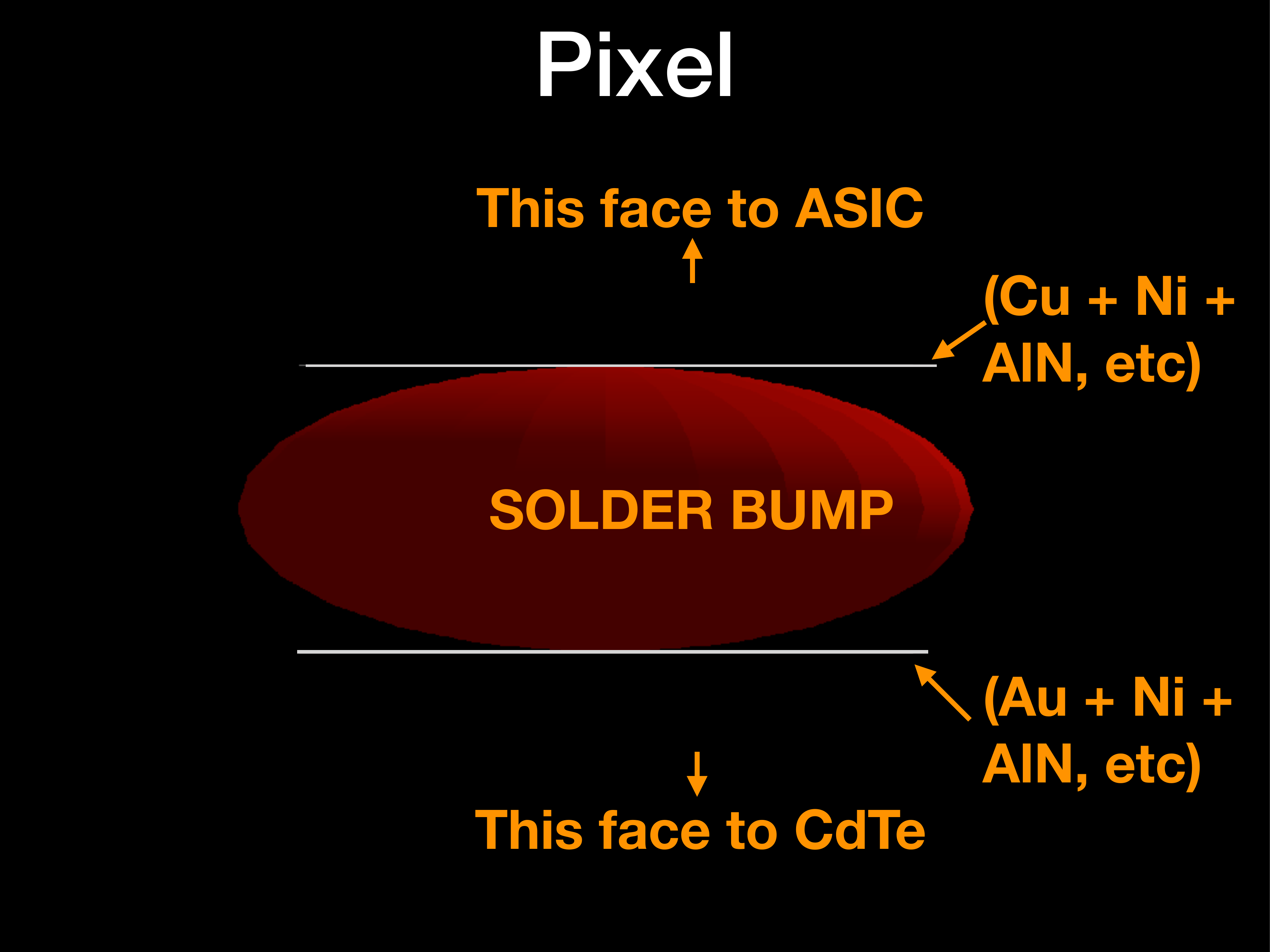}  
\end{center}
\caption{The detector geometry including sensitive material (CdTe), pixels, and ASIC. An expanded view of a pixel with  metallization layers and bump bonds is shown on the bottom right plot.
}

\label{fig:detector}
\end{figure}

\begin{figure}[tbp!]
\begin{center}
\includegraphics[width=0.8\textwidth]{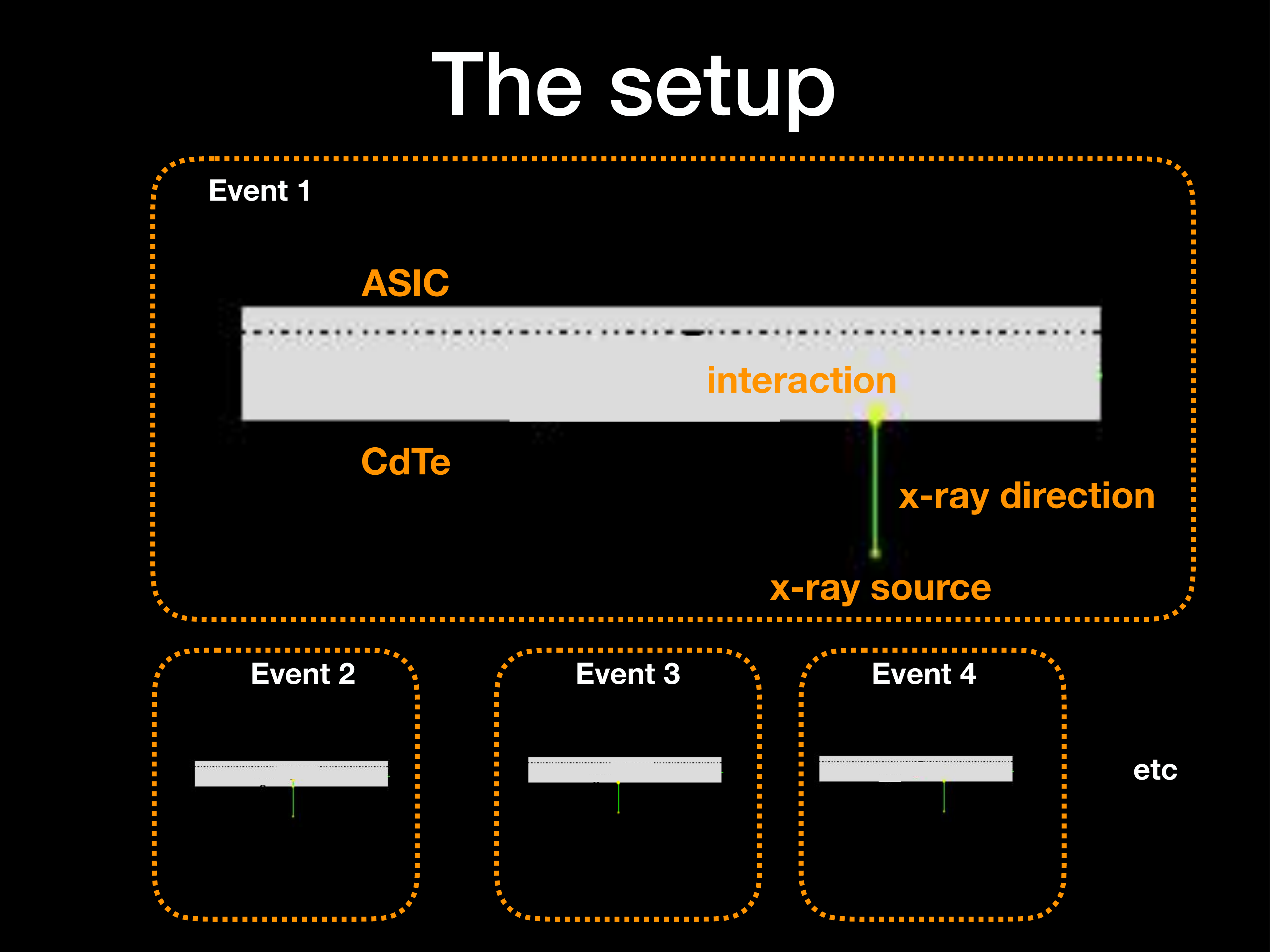}
\end{center}
\caption{Sketch of typical X-ray events as displayed by the GEANT4 OpenGL visualization. The detector is shown in gray, including sensitive material (CdTe), pixels, and ASIC. X-ray direction is indicated with a green line ending with a ``cloud'' corresponding to a photon interaction/absorption. 
}
\label{fig:source}
\end{figure}

\begin{figure}[tbp!]
\begin{center}
\includegraphics[width=0.79\textwidth]{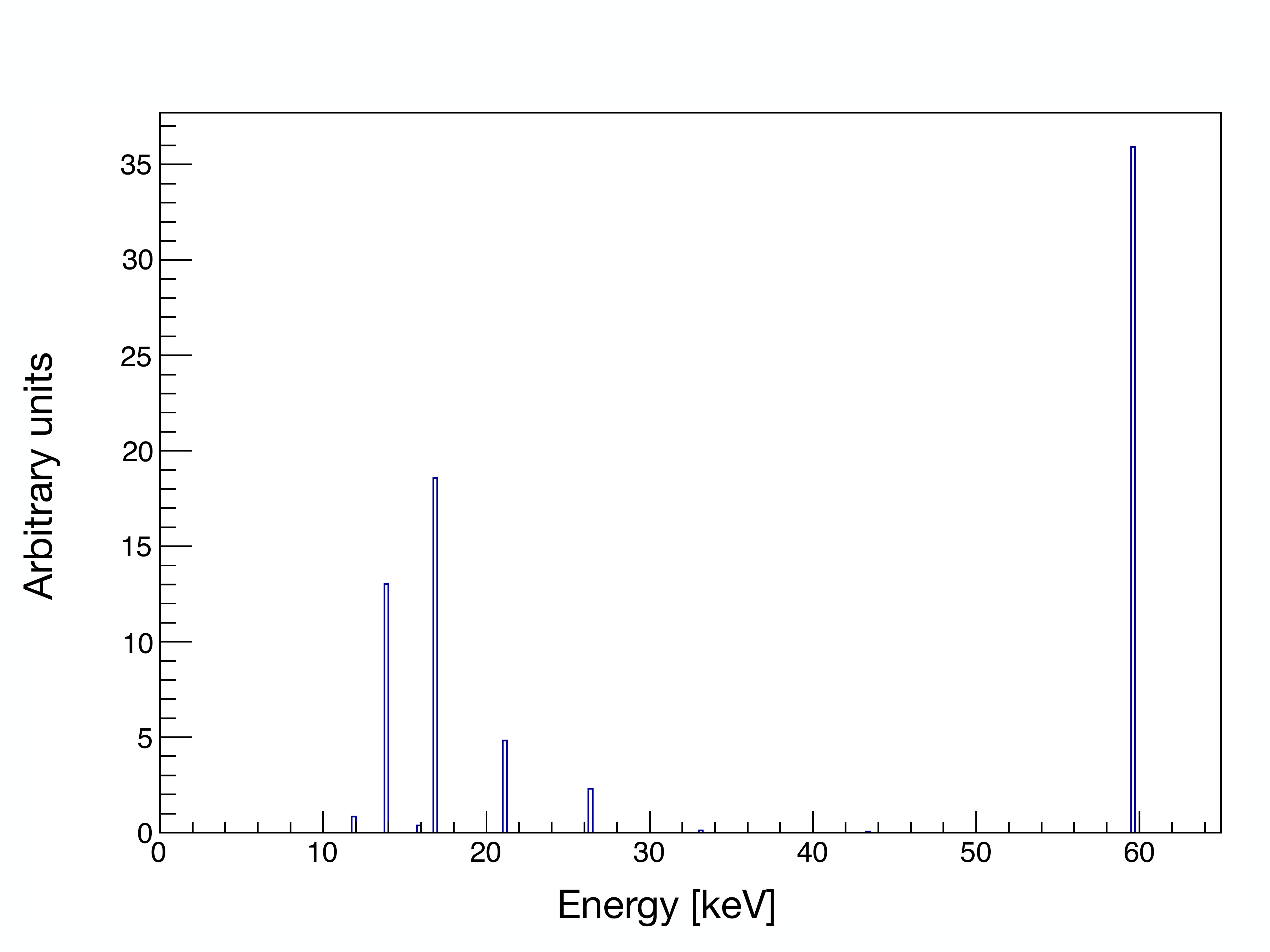}
\end{center}
\caption{Example of source energy spectrum used in the simulation for the $\mathrm{^{241}Am}$. 
}
\label{fig:energy_spectrum_W}
\end{figure}

The simulation of X-ray propagation in the detector is implemented as a C++ code within the GEANT4 framework. A typical workflow  of this code consists of the following steps. First, the detector geometry is implemented in the GDML format\footnote{Geometry Description Markup Language: http://gdml.web.cern.ch/GDML/doc/GDMLmanual.pdf}, 
as illustrated in Figure~\ref{fig:detector}. While not seen in the figure, both ASIC- and CdTe-side metallization parts include multiple layers of metals (e.g. Cu, Ni, Au, and so forth) and some aluminium nitride (AlN) coating on top. An  opening of 50~\micron~for bump bond is implemented in the AlN coating on both sides. 
Then, the particle source is placed at some distance in front of the detector, emitting X-rays in the direction perpendicular to the detector. A single X-ray photon emitted by the source is called hereafter an ``event''. The position of the source at each event is generated randomly within a plane parallel to the detector surface, to ensure uniform irradiation of the detector at different points, (Figure~\ref{fig:source}). The source energy spectrum is chosen depending on the exact task, e.g. the X-ray component of the $^{241}$Am decay spectrum shown in Figure~\ref{fig:energy_spectrum_W}, and implemented through the \mbox{\code{G4GeneralParticleSource}} class.

The physical models describing particle interactions in GEANT4 (hereafter referred to as physics lists) are versatile and are usually chosen based on the characteristic energy range and type of considered physics (electromagnetic, hadronic, thermal nuclear, etc.). 
We started with either \mbox{\code{emstandard\_opt4}} or \mbox{\code{emlivermore}} physics lists which are suggested in the GEANT4 examples for low-energy electromagnetic simulations~\footnote{https://geant4.web.cern.ch/collaboration/working\_groups/electromagnetic/}.
It was later found that these physics lists suffer from over-simplified simulation of fluorescence photons and Auger electrons. 
Namely, the range cut-off for secondary particle production by electrons and photons is set by default to 1~mm, which is then internally converted to energy for individual materials\footnote{https://geant4-userdoc.web.cern.ch/UsersGuides/ForApplicationDeveloper/fo/BookForApplicationDevelopers.pdf}.
Below this threshold, an interaction is treated as a point-like energy release. As a result, escape photons and sharing of signal between multiple pixels are not simulated to a sufficient level of accuracy in this case. In particular, the number of events triggering more than one pixel is about 4\%, which is much lower than the approximately 36\% as measured with comparable real detectors~\mbox{\citep{VEALE2014218}}.

\begin{figure}[htb!]
\begin{center}
\includegraphics[width=0.79\textwidth]{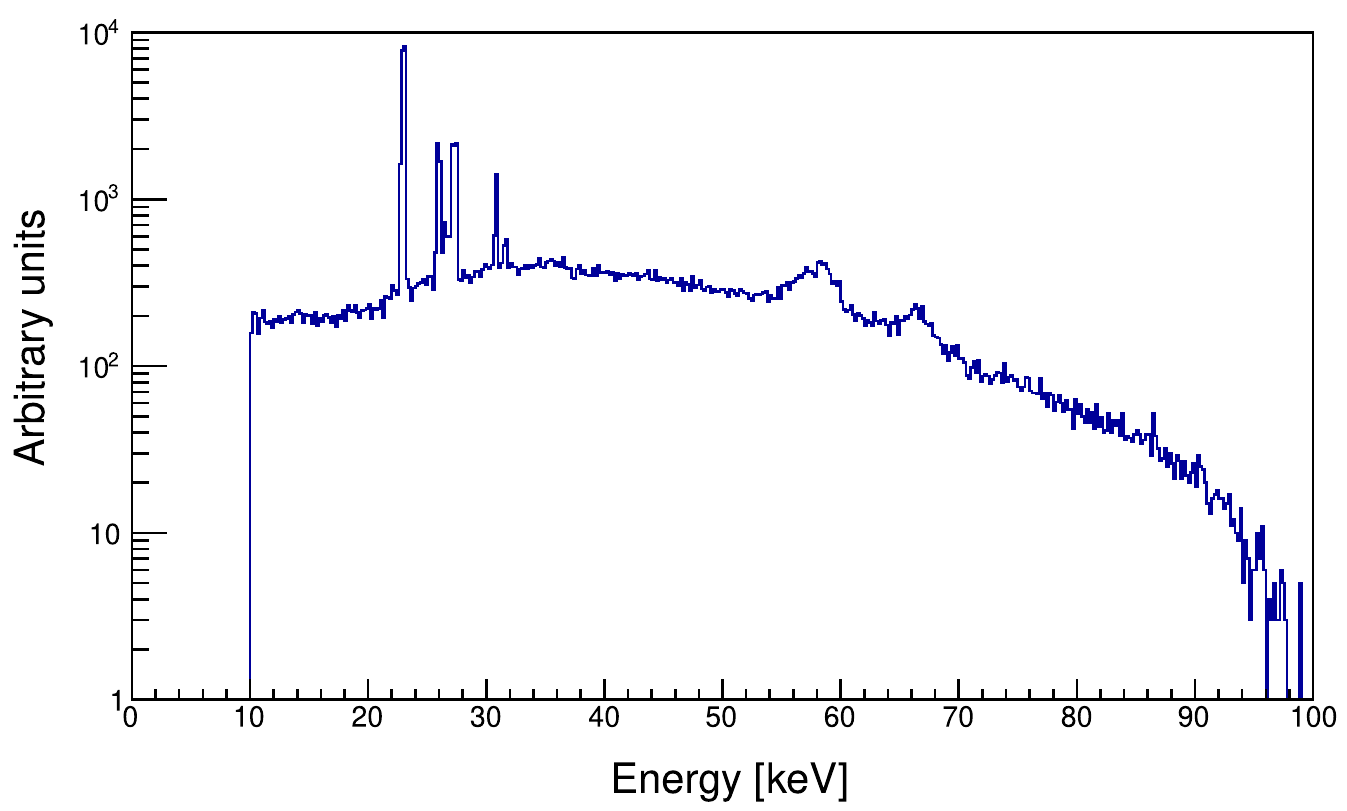}
\end{center}
\caption{An example distribution of released energy per pixel in the simulation, for a tungsten X-ray tube spectrum at 100~kVp. 
A 10~keV threshold filter is applied to the simulated detector signal to reduce the noise.}
\label{fig:fluo}
\end{figure}

In the following simulations, we have modified the \mbox{\code{emlivermore}} physics list, reducing the range cut-off for secondary production, as will be shown in the example below. In that case, the fluorescence photons and Auger electrons are simulated more accurately.
 Figure~\ref{fig:fluo} shows a typical spectrum of energy per pixel induced by a tungsten X-ray tube at 100~kVp, featuring characteristic peaks between 20 and 30~keV similar to other dedicated simulations elsewhere~\citep{Xu2011EvaluationOE}. Note that such emission lines are not present if the standard physics list is used. The number of events triggering more than one pixel can be up to 50\% with the modified physics list, which is more realistic for this type of detector and pixel size~\citep{Krzyzanowska:2018}.
We, therefore, use the modified physics list as a baseline in our studies. As a result of the GEANT4-based simulation described here,  a per-event 3D map of energy releases in the detector can be produced. 

The latter simulation can be reproduced by following the GEANT4 example~\code{TestEm5}\footnote{Located in \texttt{\$G4INSTALL/../examples/extended/electromagnetic/TestEm5}},  using  the physics list from \code{pixe.mac} macro file of the example. A standard GEANT4 installation is required (version \code{4.10.04.p02} or higher) with no additional modules needed. The following commands have to be present in the macro:\\

\noindent\texttt{/testem/phys/addPhysics emlivermore\\
/run/setCut 0.01 mm\\
/run/setCutForAGivenParticle gamma 0.5 um\\
/process/em/deexcitationIgnoreCut true\\
/process/em/fluo true\\
/process/em/pixe true\\
/process/em/auger true\\
/process/em/augerCascade true\\
}

\section{Simulation of detector response}
\label{sec:response}

In this section, we describe the procedure for converting the released energy into the detector signal, hereafter referred to as the~\emph{detector response simulation}. To validate our approach, we compare the simulation results with data from the real detector described in~\cite{DTSeamlessTile} (1~mm thickness CdTe, 150~\micron~pixel pitch). The detector response simulation consists of the following components:
\begin{itemize}
\item ASIC signal non-linearity
\item Charge sharing
\item Energy resolution effects
\item Electronic noise
\end{itemize}

\subsection{ASIC signal non-linearity}

In order to check the detector's signal linearity, the developers of~\cite{DTSeamlessTile}
measured the X-ray fluorescence (XRF) spectra for five different materials: Mo, Ag,
Gd, W, and Pb. The location of  XRF spectral lines is well known for these
materials and may be found e.g. in the NIST database\footnote{National Institute of Standards and Technology (NIST), X-ray transition energies by element(s), transition(s), and/or energy/wavelength range: https://physics.nist.gov/PhysRefData/XrayTrans/Html/search.html}. 
The most important
are $K_\alpha$ and $K_\beta$ lines. Figure~15 in~\citep{DTSeamlessTile} shows
the relevant parts of these spectra as measured by 
the authors. Figure~16
in~\cite{DTSeamlessTile} establishes the connection between the least sensitive bits (LSBs) and
photon energies for all the observed peaks. 
As is seen from the same figure, 
the 
detector's energy response is extremely close to linear for the photon
flux used in their measurements. This fact removes the necessity to use any non-linear expression to convert the detector's scale to the true one, and determine its parameters. A simple relation may be used instead~\cite[Equation~(6)]{DTSeamlessTile}. Note, that this holds true only for the low flux used by the authors of~\citep{DTSeamlessTile}:
\begin{equation}
E ~[keV] = 0.386 \cdot \text{\textit{threshold}} ~[LSB] - 7.236
\label{eq-Prototype-energy-scale-paper}
\end{equation}

\subsection{Energy resolution and electronic noise}

To model the electronic noise and energy resolution of the detector, we ``smear'' the energy response of ASIC in each X-ray event, so that the observed energy is determined as follows:

\begin{equation}
E_{\mathrm{obs}} = E_{\mathrm{dep}}\left( 1 +  \sigma_E \cdot Rand_1\right) + \sigma_{\mathrm{nosie}} \cdot Rand_2
\label{eq:easic}
\end{equation}

\noindent
where $Rand_1$ and $Rand_2$ are two independent random variables generated according to a Gaussian distribution with $mean=0$ and $\sigma=1$. Note that $\sigma_E$ is dimensionless, and $\sigma_{noise}$ has the units of energy. Their values will be determined later in Section~\mbox{\ref{sec-model-parameters}}.

The 59.5~keV peak of $^{241}$Am decay radiation is 
useful for determining 
the 
detector's energy resolution. According to~\cite{DTSeamlessTile}, the
full width at half maximum (FWHM\footnote{For Gaussian distribution FWHM$=2\sqrt{2ln(2)}\sigma\approx2.355\sigma$}) of this line is 9~LSB, 
which corresponds to the value of 
$\Delta E \approx 3.48$~keV
, or the energy resolution:
\begin{equation}
\eta = \frac{\Delta E}{E} \cdot 100 \% = \frac{3.48~keV}{59.5~keV} \cdot 100 \%  \approx 6 \% 
\end{equation}

This value of $\Delta E$ includes both the actual energy resolution
of the crystal itself and the ASIC noise level. While the absolute value of
the energy resolution of the crystal is proportional to the photon energy, the ASIC
noise level is supposed to be constant for all energies.
So the values of $\sigma_E$ and $\sigma_{noise}$ may be determined by fitting the $^{241}$Am peak in the simulated detector response to the experimentally measured one (see Figure~17 in~\cite{DTSeamlessTile}), as will be shown further in this section.

\subsection{Charge sharing}

As noted in Section~\mbox{\ref{sec:geant4}}, the inclusion of fluorescence photons in the simulation increases the number of events with more than one triggered pixel from about 4\% to as high as 50\% (the exact fraction depends on the pixel pitch). Therefore, the absolute majority of signal sharing between different pixels in the detector is attributed to the fluorescence photons, simulated in the GEANT4 part. Nonetheless, signal sharing due to charge diffusion is not negligible and must be taken into account as well. We employ a simple Gaussian model following the prescription~\citep{4437135,KIM2011233} to simulate the charge diffusion. First, the deposited energy in the sensitive detector is logically split into a sufficiently high number of equal parts. Then, each part is randomly moved within the detector plane from the initial point according to a 2-dimensional Gaussian distribution with
standard deviation $\sigma_c$, which reasonably approximates the effect of charge sharing on to the charge carriers within the electrical field of the detector crystal.

A value of $\sigma_c=11$~\micron~was used as a starting value, estimated according to~\citep{4437135} for the detector of 1~mm thickness and 400~V bias voltage.

\subsection{Fitting the model parameters}

\label{sec-model-parameters}

As noted previously, a very important measurement presented in~\cite{DTSeamlessTile} is the~$^{241}$Am X-ray emission spectrum (Figure~17 in~\cite{DTSeamlessTile}).
It has a distinct solitary line at $59.5$~keV 
and allows the estimate of the detector response parameters, as described below.

\begin{figure}[htb!]
\begin{center}
\includegraphics[width=9cm]{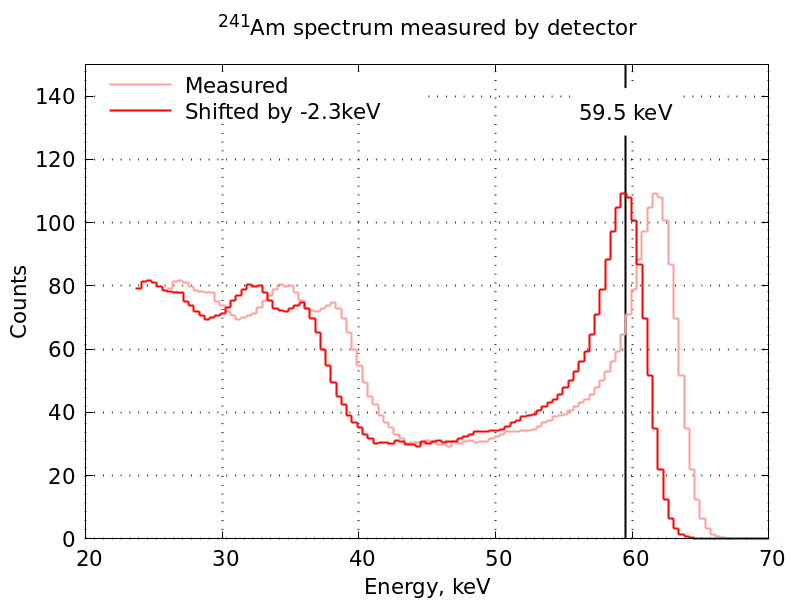}
\end{center}
\caption{The energy shift required to match the location of the major
         $^{241}$Am peak with the well-known value~\cite{DTSeamlessTile}.
         }
\label{fig-Prototype-energy-shift}
\end{figure}

\emph{Energy shift}. As is seen in 
Figure~17 in~\cite{DTSeamlessTile},
this peak is located at
178~LSB, which, according to Equation~\ref{eq-Prototype-energy-scale-paper},
corresponds to about 61.5~keV, while the true $^{241}$Am peak is located at
about 59.5~keV. Therefore, it is assumed that the whole measured spectrum is shifted by
about $2$~keV to higher energies (Figure~\ref{fig-Prototype-energy-shift}). 
This assumption will be justified below by comparison of the simulated and measured spectra.

\begin{figure}[htb!]
\begin{center}

\includegraphics[width=9cm]{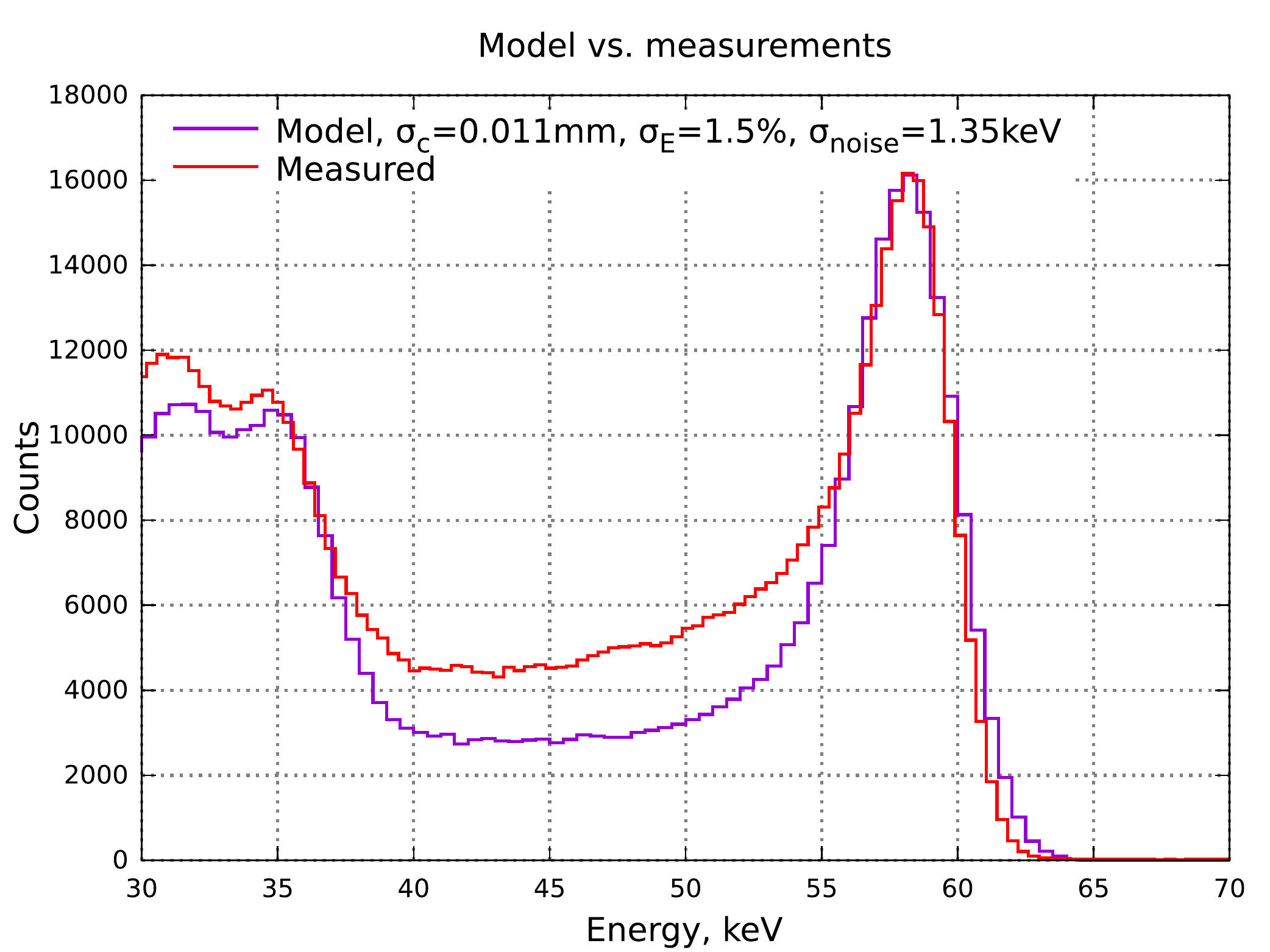}
\end{center}
\caption{
Simulated response to $^{241}$Am decay spectrum (violet line) compared to the actually measured data (red line) ignoring the peak-to-valley ratio. The data points for the red plot were obtained directly from the authors of~\mbox{\cite{DTSeamlessTile}}.}
\label{fig-Prototype-response-first}
\end{figure}

Having set the starting values, we run the simulation using the $^{241}$Am decay spectrum (Figure~\ref{fig:energy_spectrum_W}) as an input. This way we obtain the simulated detector response and compare it to the experimentally measured one (Figure~17 in~\cite{DTSeamlessTile}).
Next, we tune the $\sigma_E$ and $\sigma_{noise}$ parameters (energy resolution and noise level) during the  
detector simulation in order to match the measured
response to the $^{241}$Am 59.5~keV photons. 
The result is presented in Figure~\ref{fig-Prototype-response-first} where the energy resolution is chosen to resemble the measured peak width while ignoring the peak-to-valley ratio. It was achieved with 
$\sigma_E = 1.5\%$ and $\sigma_{noise} = 1.35$~keV. In fact, these values correspond to a root mean square (RMS) of the observed energy (Equation~\ref{eq:easic}) at the  $^{241}$Am peak of about 1.62~keV and respectively the FWHM of about 3.81~keV,  which is very close to the experimentally determined effective energy resolution of the $^{241}$Am peak, 3.48~keV~\cite{DTSeamlessTile}.

\begin{figure}[htb!]
\begin{center}
\includegraphics[width=9cm]{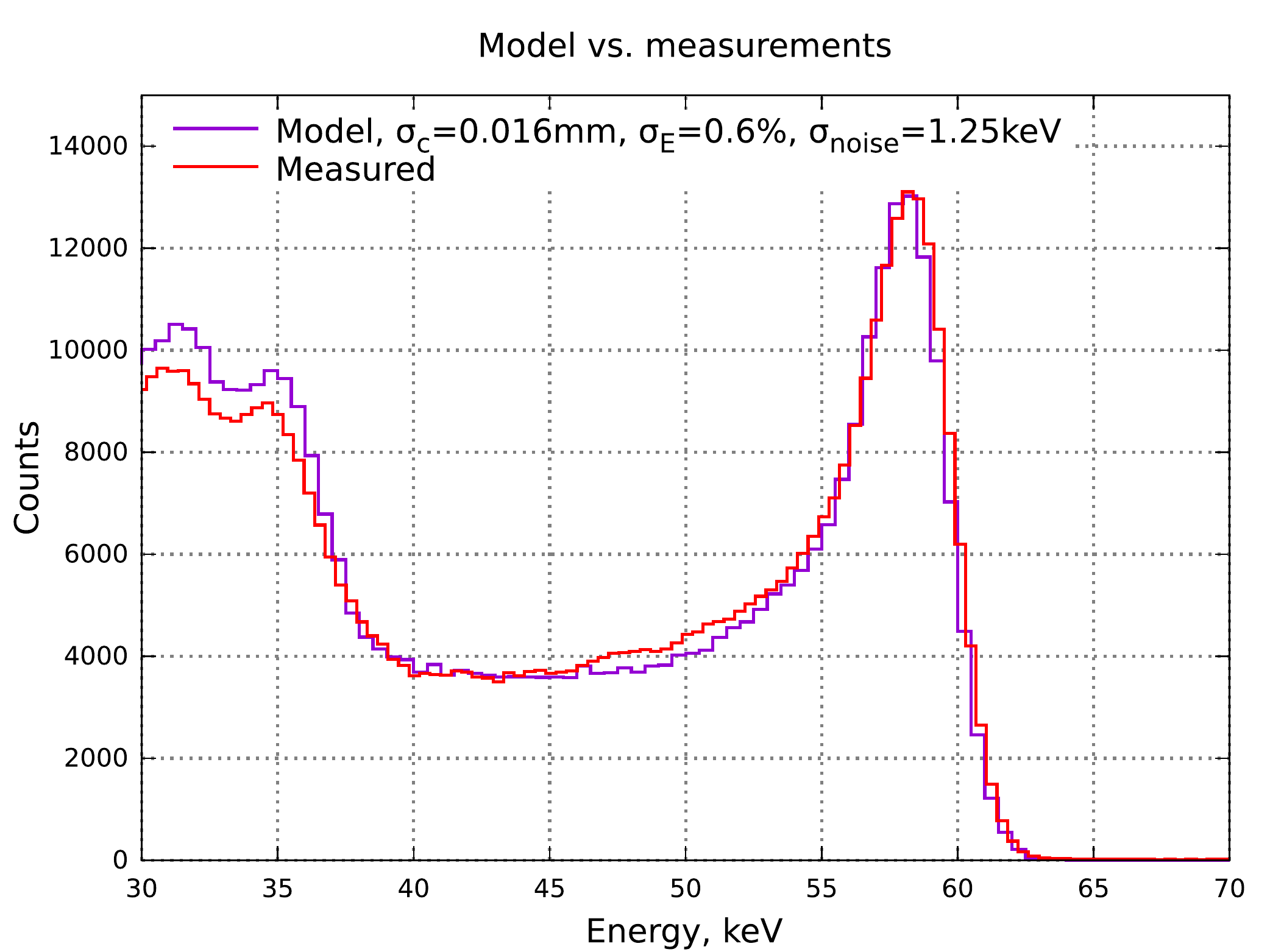}
\end{center}
\caption{Simulated response to $^{241}$Am decay spectrum (violet
         line) compared to the actually measured data (red line). A much better
         correspondence of the peak-to-valley ratio is achieved here. Note that
         this also required a certain correction of the energy resolution.}
\label{fig-Prototype-response-second}
\end{figure}

\emph{Peak-to-valley ratio.} Despite a very good correspondence between the simulated and measured data around the $^{241}$Am peak in Figure~\ref{fig-Prototype-response-first}, 
an evident discrepancy between the 
curves can be seen in the 
peak-to-valley ratio. 
According to \cite{DTSeamlessTile}, the peak-to-valley ratio of 
the real
detector
is about~4, while our current model shows a value of about~6. 
Tuning the $\sigma_c$ parameter
for 
our 
detector model allowed to obtain a correct peak-to-valley ratio and
achieve a better match between the simulated detector response and the measured
one. The result is presented in Figure~\ref{fig-Prototype-response-second}.
Note that the adjustment of the charge cloud size also influences the other
parameters describing the energy resolution. For the purpose of this study, the best results are achieved
with $\sigma_c = 16$~\micron, $\sigma_E = 0.6\%$ and $\sigma_{noise} = 1.25$~keV
, which are considered as a baseline further in the paper.

\section{Spectrum restoration}
\label{sec:unfolding}

Once a good correspondence between the model and the data measured by a real detector is achieved, we 
aim 
to use our model to reconstruct the initial radiation spectrum given the detector response.

For this purpose, we are going to use the so-called ``unfolding'' approach, in which  the spectrum measured by the detector (data) is converted into the 
initial
(true) spectrum of X-rays impinging the detector using the detector response operator. This approach relies on the detector simulation model developed in Sections~\ref{sec:geant4} and~\ref{sec:response}. The spectrum unfolding principle is described below.

The spectrum measured by the detector (data) is related to the primary X-ray spectrum through the linear transformation:
\begin{gather}
 \begin{pmatrix} n_{1}^{\mathrm{data}} \\ n_{2}^{\mathrm{data}} \\ ...  \\ n_{N}^{\mathrm{data}} \end{pmatrix}
 =
  \begin{pmatrix}
   M_{11} &    M_{12}  &    ...  &   M_{1N}\\
   M_{21} &    M_{22}  &    ...  &   M_{2N}\\
   ... &    ...  &    ...  &   ... \\
   M_{N1} &    M_{N2}  &    ...  &   M_{NN}\\
   \end{pmatrix} 
   \cdot
    \begin{pmatrix} n_{1}^{\mathrm{true}} \\ n_{2}^{\mathrm{true}} \\ ...  \\ n_{N}^{\mathrm{true}} \end{pmatrix}
\end{gather}
where $n_{i}^{\mathrm{data}}$ is the number of triggered events in $i$-th energy bin,  $n_{i}^{\mathrm{true}}$ is the number of emitted X-rays in $i$-th energy bin,  $N$ is the number of energy thresholds of the photon counting detector, and $M_{ij}$ is the probability of X-ray with $j$-th energy to trigger an event in $i$-th energy bin. This can be written in matrix equation form:
\begin{equation}
\hat{n}^{\mathrm{data}} = \hat{M}  \cdot \hat{n}^{\mathrm{true}} 
\label{eq:folding}
\end{equation}
The energy response matrix, $\hat{M}$, can be estimated in the simulation. Deriving the initial spectrum from the reconstructed one is reduced to the  following inversion problem:
\begin{equation}
\hat{n}^{\mathrm{true}} = \hat{M}^{-1}  \cdot \hat{n}^{\mathrm{data}} 
\label{eq:unfolding}
\end{equation}
Note however that the inverse of $\hat{M}$ 
usually does not have an exact solution and regularization methods are required. Iterative Bayesian inference procedures are well known as the right tool for such problems~\cite{DAgostini1995,DAgostini2010}. In our case, we use a custom implementation of the Bayesian unfolding. Note  that the ``smoothing'' of the solution in case of a large number of energy thresholds (100 or more) is applied, otherwise the restored spectrum yields non-physical fluctuations.

Having a good enough software detector model with realistic parameters, one can produce a detector response matrix $\hat{M}$ to use in the spectrum restoration procedure.

It should be stressed that the response matrix which is at the core of the unfolding method depends strongly on the detector configuration, in particular material type, thickness, composition, pixel size/pitch, angular spread of impinging X-rays, etc. Hence, for each particular detector configuration, a separate matrix has to be generated in the simulation. In other words, no analytic formula can be used to adopt a model from one detector configuration to another one. However, the presented approach provides a recipe that allows the preparation of models for other CdTe pixelated detectors in different configurations.

\subsection{Restoration from simulated spectra}

In order to verify the self-consistency of our approach, in this subsection, we simulate several test cases using our tuned model and apply the spectrum restoration algorithm to these simulations. They include the monochromatic sources from 20~keV to 90~keV and a tungsten X-ray tube spectrum.

\emph{Monochromatic source}. The monochromatic source, as can be produced by synchrotron facilities, appears to be the easiest task for the spectrum restoration algorithm, as it consists of a solitary sharp spectral line that usually remains recognizable even after the application of all kinds of radiation interaction and electronics effects described in Section~\ref{sec:response}. For example, as can be seen in Figures~\ref{fig-Prototype-response-first} and~\ref{fig-Prototype-response-second}, the detector response to the $^{241}$Am decay spectrum demonstrates a clearly distinguishable 59.5~keV peak.

\begin{figure}[htbp!]
\begin{center}
  \begin{minipage}[c]{0.48\linewidth}
  \includegraphics[width=\linewidth]{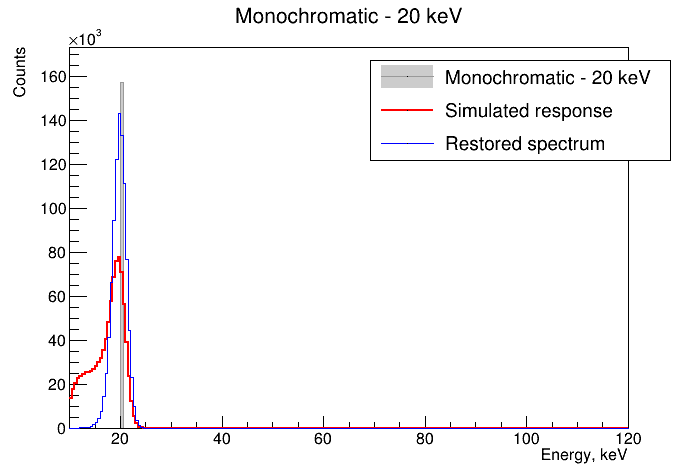}
  \end{minipage}
  \hfill
  \begin{minipage}[c]{0.48\linewidth}
  \includegraphics[width=\linewidth]{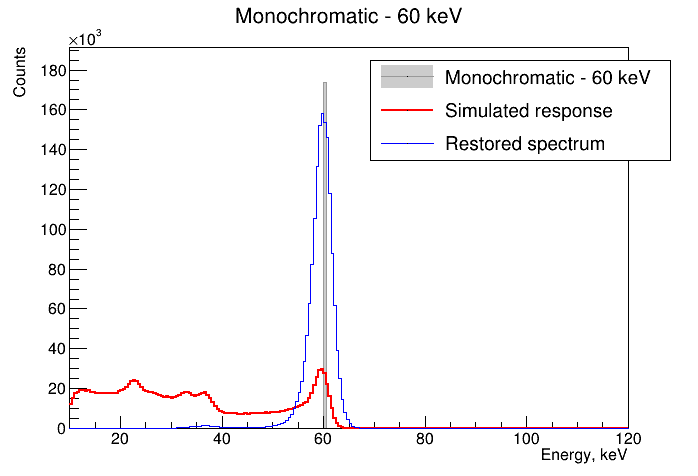}
  \end{minipage}
  \\
  \begin{minipage}[c]{0.48\linewidth}
  \includegraphics[width=\linewidth]{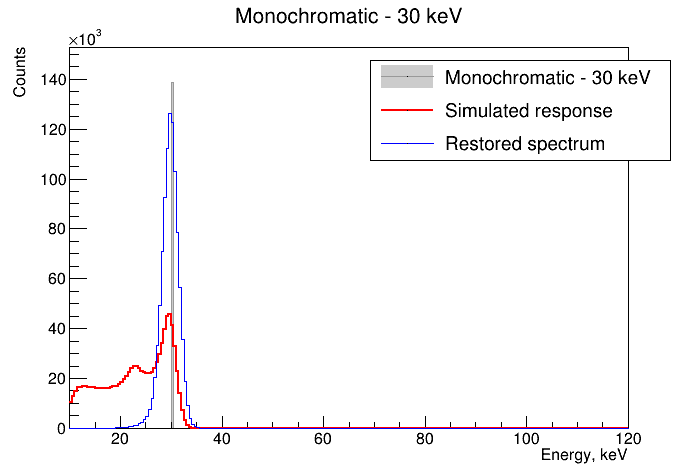}
  \end{minipage}
  \hfill
  \begin{minipage}[c]{0.48\linewidth}
  \includegraphics[width=\linewidth]{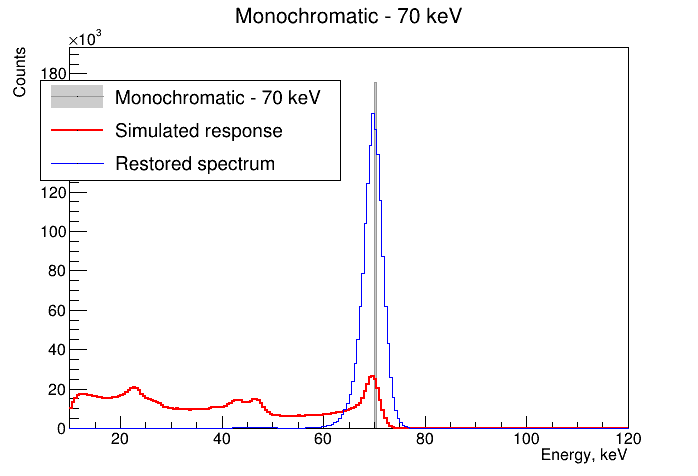}
  \end{minipage}
  \\
  \begin{minipage}[c]{0.48\linewidth}
  \includegraphics[width=\linewidth]{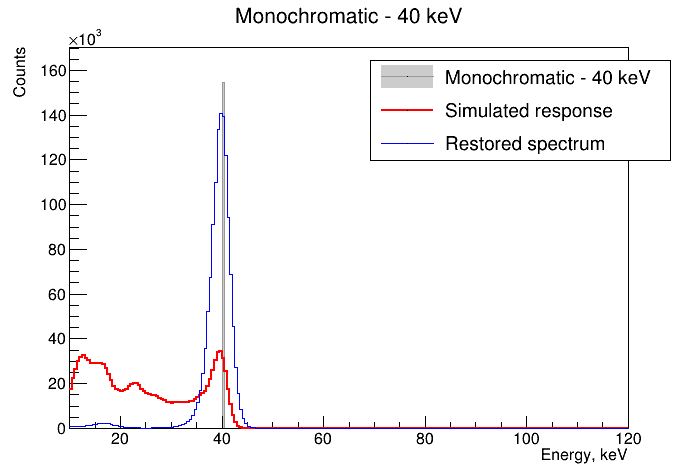}
  \end{minipage}
  \hfill
  \begin{minipage}[c]{0.48\linewidth}
  \includegraphics[width=\linewidth]{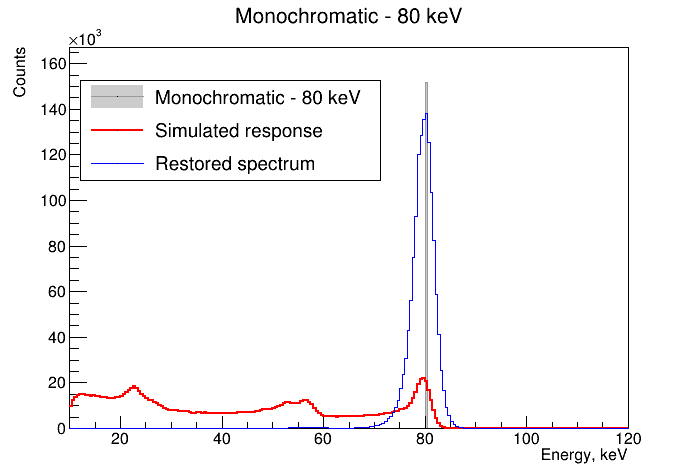}
  \end{minipage}
  \\
  \begin{minipage}[c]{0.48\linewidth}
  \includegraphics[width=\linewidth]{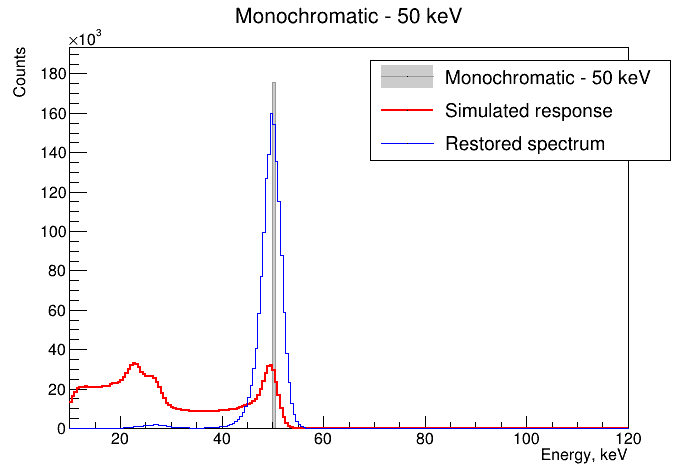}
  \end{minipage}
  \hfill
  \begin{minipage}[c]{0.48\linewidth}
  \includegraphics[width=\linewidth]{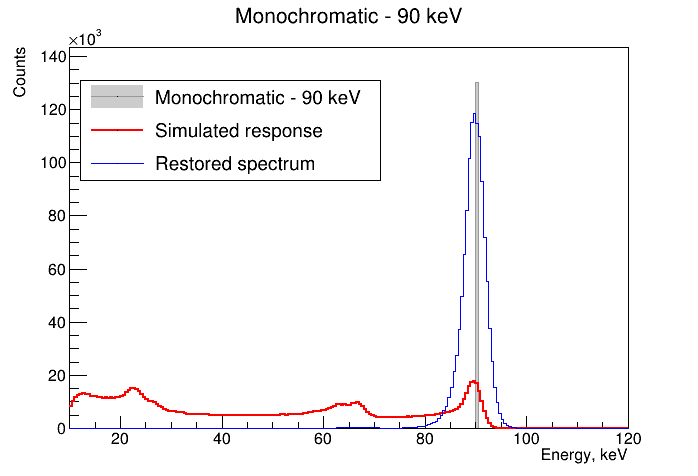}
  \end{minipage}
\end{center}
\caption{Monochromatic spectra restoration for simulated data. Narrow grey bars mark
         the location of the original monochromatic spectral lines.}
\label{fig-Prototype-unfolding-mono}
\end{figure}

The results of the detector response simulation together with the restored initial spectra are presented in Figure~\ref{fig-Prototype-unfolding-mono}, for monochromatic X-ray spectral lines between 20 and 90~keV, with a step of 10~keV. 
As can be seen from these plots, the
monochromatic spectra are reconstructed from the simulated detector response
very accurately. It is important to note that in spite of the exact shape of the initial spectrum being known \textit{a priori}, this information is neither used for evaluating the detector response matrix $\hat{M}$, nor for solving or regularising the Equation~(\ref{eq:unfolding}). In other words, we do not imply any prior guess on the solution.

\FloatBarrier

\begin{figure}[htbp!]
\begin{center}
\includegraphics[width=13cm]{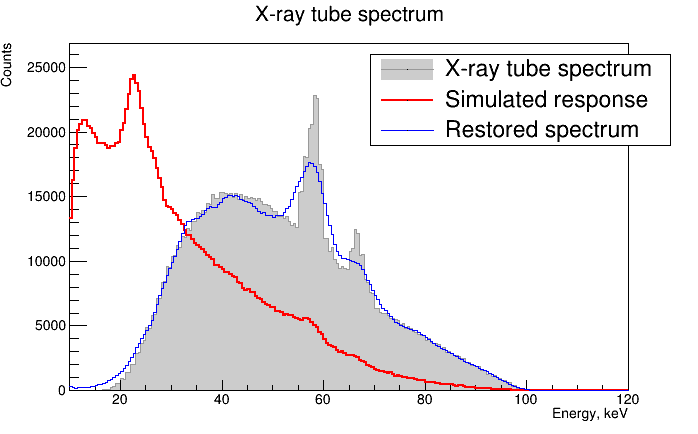}
\end{center}
\caption{Spectrum restoration based on the simulated response of the
         detector to the radiation of a tungsten X-ray tube. The grey shaded area shows the initial spectrum used to simulate the detector response (generated by SpekPy software~\mbox{\cite{SpekPy}}), the red curve shows the detector response itself, and the blue curve shows the restored spectrum based on the detector response (red curve).}
\label{fig-Prototype-unfolding-XRayTube}
\end{figure}

\emph{X-ray tube}. The success achieved with monochromatic spectra restoration above does not necessarily guarantee 
 the possibility of solving more complex scenarios. In particular, a tungsten X-ray tube spectrum contains a series of sharp peaks of the characteristic radiation on top of a smooth continuous background of bremsstrahlung radiation. Using the X-ray tube test case, we can assess the efficiency of our spectrum restoration algorithm for more complex spectra encountered in the real world. We use the same detector response model as before and solve Equation~(\ref{eq:unfolding}) employing the same $\hat{M}$ matrix, but taking the simulated detector response to the tungsten X-ray tube at 100~kVp as observed data. 
The true X-ray tube spectrum (grey shaded area), the simulated detector response (red line), and the restored spectrum (blue line) are shown in Figure~\ref{fig-Prototype-unfolding-XRayTube}.


As can be seen from Figure~\ref{fig-Prototype-unfolding-XRayTube}, the X-ray tube spectrum is also reconstructed
with high accuracy. Both characteristic tungsten peaks are reproduced in the restored spectrum, although not as sharp as in the input X-ray tube spectrum. The shape of the reconstructed bremsstrahlung continuous spectrum matches very well the original one. 
Again, we emphasize that no prior knowledge of the true spectrum is used at the stage of spectrum restoration. Aside from the detector response model $\hat{M}$, the only input to the spectrum restoration algorithm is the simulated detector response (red curve in Figure~\ref{fig-Prototype-unfolding-XRayTube}).

\FloatBarrier

\subsection{Restoration from measured spectra}

\begin{figure}[htbp!]
\begin{center}
\includegraphics[width=13.8cm]{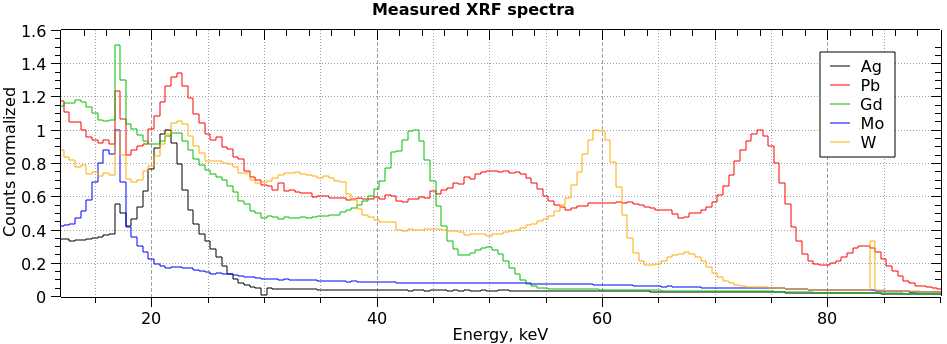}
\end{center}
\caption{XRF spectra for Mo, Ag, Gd, W, and Pb measured with the detector from~\cite{DTSeamlessTile}.
}
\label{fig-Prototype-measured-XRF}
\end{figure}

After the simulation-based spectrum restoration was performed
successfully, we applied the same procedure and detector response model to the experimental data measured by a real detector.
We used the XRF spectra
for several targets: Mo, Ag, Gd, W, and Pb from~\cite{DTSeamlessTile} as an input (Figure~\ref{fig-Prototype-measured-XRF}).
Each spectrum in Figure~\mbox{\ref{fig-Prototype-measured-XRF}} is normalized by the height of the respective XRF peak.

\begin{figure}[htbp!]
\begin{center}
\includegraphics[width=8cm]{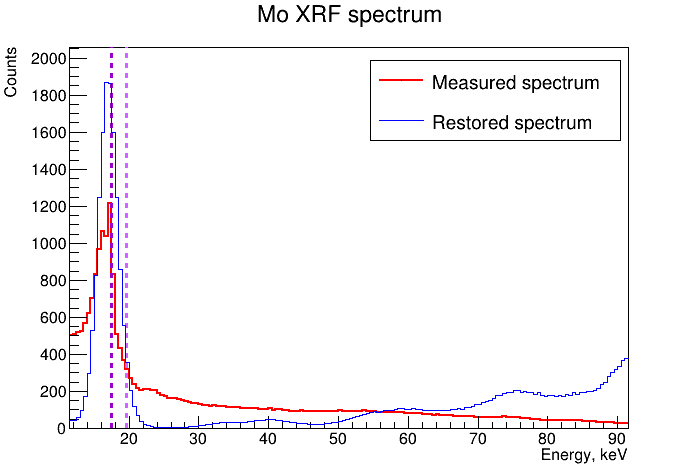}
\end{center}
\caption{XRF spectrum restoration for Mo target. Dashed violet lines mark the
K$_\alpha$ and K$_\beta$ fluorescence lines of molybdenum.}
\label{fig-Prototype-unfolding-XRF-Mo}
\end{figure}

\begin{figure}[htbp!]
\begin{center}
\includegraphics[width=8cm]{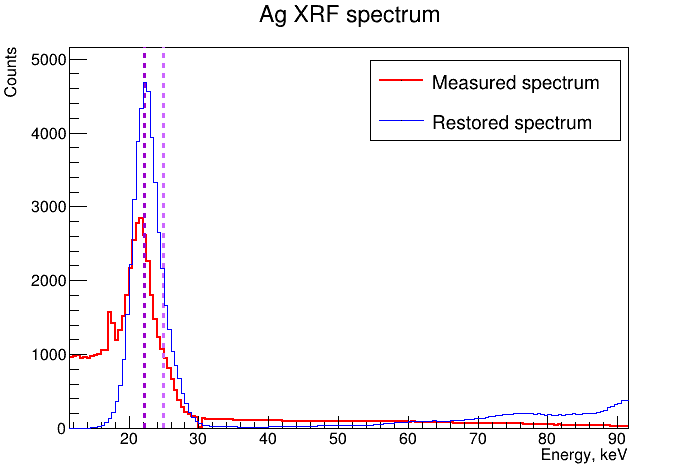}
\end{center}
\caption{XRF spectrum restoration for Ag target. Dashed violet lines mark the
K$_\alpha$ and K$_\beta$ fluorescence lines of silver.}
\label{fig-Prototype-unfolding-XRF-Ag}
\end{figure}

\begin{figure}[htbp!]
\begin{center}
\includegraphics[width=8cm]{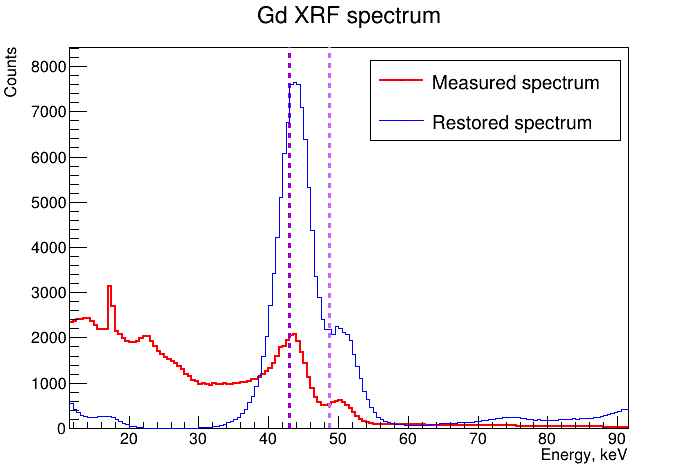}
\end{center}
\caption{XRF spectrum restoration for Gd target. Dashed violet lines mark the
K$_\alpha$ and K$_\beta$ fluorescence lines of gadolinium.}
\label{fig-Prototype-unfolding-XRF-Gd}
\end{figure}

\begin{figure}[htbp!]
\begin{center}
\includegraphics[width=8cm]{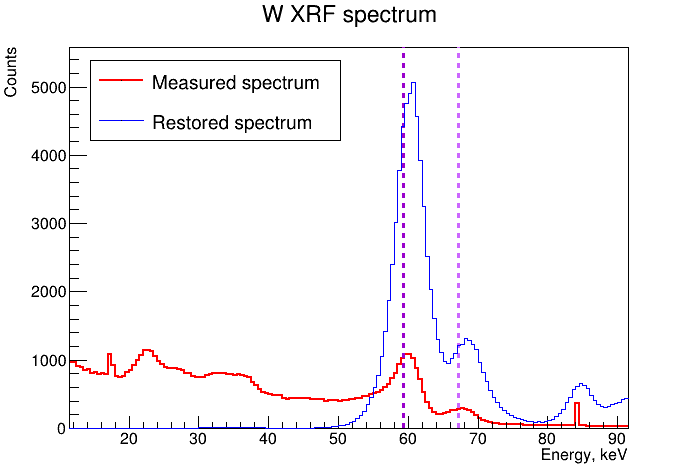}
\end{center}
\caption{XRF spectrum restoration for W target. Dashed violet lines mark the
K$_\alpha$ and K$_\beta$ fluorescence lines of tungsten.}
\label{fig-Prototype-unfolding-XRF-W}
\end{figure}

\begin{figure}[htbp!]
\begin{center}
\includegraphics[width=8cm]{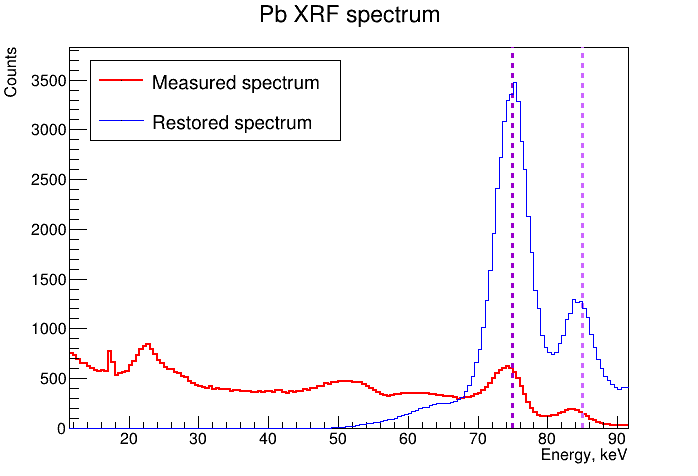}
\end{center}
\caption{XRF spectrum restoration for Pb target. Dashed violet lines mark the
K$_\alpha$ and K$_\beta$ fluorescence lines of lead.}
\label{fig-Prototype-unfolding-XRF-Pb}
\end{figure}

The results of the XRF spectrum restoration based on these measurements
are presented in 
Figures~\ref{fig-Prototype-unfolding-XRF-Mo}-\ref{fig-Prototype-unfolding-XRF-Pb}.
As can be seen from the figures, 
the
reconstructed XRF spectra (blue lines) demonstrate good agreement with the
expected locations of the K$_\alpha$ and K$_\beta$ lines. Both lines can be seen
distinctly in Gd, W, and Pb spectra.
Noteworthy, the spectrum restoration algorithm also reconstructs minor
artifacts in the input signal like single-bin peaks in Gd and W measurements, despite not being trained on those.

It is also interesting that some high-energy tails can be observed for the measured XRF spectra of
Mo, Ag, and Gd (most visible in Figure~\ref{fig-Prototype-unfolding-XRF-Mo}). These can be caused by the hardware signal processing of the detector and lead to the artifact of a rising baseline of the reconstructed spectrum at the high energy end (blue line in Figures~\ref{fig-Prototype-unfolding-XRF-Mo}-\ref{fig-Prototype-unfolding-XRF-W}). 
By construction, the algorithm predicts what looks like the most probable
input spectrum that could produce such a tail, however creating non-physical results in the process, as the detector model has not been designed to incorporate such a feature of the high energy tail. 
Nonetheless, as it was demonstrated above
with the monochromatic spectra tests 
(Figure~\mbox{\ref{fig-Prototype-unfolding-mono}}), this does not happen in cases 
when the measurements do not contain the high-energy tail. 
We intend to address this artifact and the correction thereof in a subsequent study 
and perform per-pixel spectrum restoration profiting from both spectral and spatial information~\citep{DAgostini1995}. 

\FloatBarrier

\section{Conclusions}
\label{sec:conclusions}

A spectrum restoration algorithm has been developed that is based on a simulated model of a pixelized CdTe photon-counting X-ray detector (CdTe). It includes the simulation of X-ray propagation in the detector material, as well as the production and propagation of secondary particles and photons. In addition, the modeling of the signal response of the detector is developed, including the charge correction, electronic noise, energy resolution, gain non-linearity and signal clustering algorithm. The calibration of the model is performed using the real data of $\mathrm{^{241}Am}$  
source measured with a photon counting detector developed by Kauppinen et al.~\cite{DTSeamlessTile}. 

Based on the obtained detector model, we applied the spectrum restoration algorithm to the simulated detector response for various test signals and obtained a good correspondence between the ground truth and restored spectra. In particular, a set of monochromatic input spectra as well as the tungsten X-ray tube spectrum at 100~kVp have been restored with good accuracy.

In the next step, we applied the same detector model together with the same spectrum restoration procedure to the experimentally measured XRF spectra. This combination of the detector model and spectrum restoration algorithm produced excellent results revealing the narrow spectral lines close to the known locations of the characteristic K$_\alpha$ and K$_\beta$ lines for the respective target materials.
The restoration of the XRF spectra also revealed a limitation of the model in its current form: by producing a nonphysical artifact in the restored spectra. We intend to investigate and address this artifact in  future research on this topic.

Finally, the developed model-based spectrum restoration technique is a promising method to address the charge sharing problem of photon-counting detectors with small pixel sizes, that suffer from this effect.

\section{Acknowledgements}

The authors acknowledge the support of Detection Technology PLC, Finland. Ukrainian authors, A. Tykhonov and V. Smoliar are indebted to the resilience and courage of the Armed Forces of Ukraine, for keeping their loved ones safe during the course of this work. 

\bibliographystyle{elsarticle-num}
\bibliography{bibliography}

\begin{thebibliography}{10}
\expandafter\ifx\csname url\endcsname\relax
  \def\url#1{\texttt{#1}}\fi
\expandafter\ifx\csname urlprefix\endcsname\relax\def\urlprefix{URL }\fi
\expandafter\ifx\csname href\endcsname\relax
  \def\href#1#2{#2} \def\path#1{#1}\fi

\bibitem{Maj_2012}
P.~Maj, A.~Baumbaugh, G.~Deptuch, P.~Grybos, R.~Szczygiel, Algorithms for
  minimization of charge sharing effects in a hybrid pixel detector taking into
  account hardware limitations in deep submicron technology, Journal of
  Instrumentation 7~(12) (2012) C12020.
\newblock \href {https://doi.org/10.1088/1748-0221/7/12/c12020}
  {\path{doi:10.1088/1748-0221/7/12/c12020}}.

\bibitem{Ullberg_2013}
C.~Ullberg, M.~Urech, N.~Weber, A.~Engman, A.~Redz, F.~Henckel, Measurements of
  a dual-energy fast photon counting cdte detector with integrated charge
  sharing correction, Medical Imaging 2013: Physics of Medical Imaging (2013).
\newblock \href {https://doi.org/10.1117/12.2007892}
  {\path{doi:10.1117/12.2007892}}.

\bibitem{geant4}
S.~Agostinelli, et~al., Geant4 -- a simulation toolkit, Nucl. Instrum. Meth.
  506~(3) (2003) 250 -- 303.
\newblock \href {https://doi.org/10.1016/S0168-9002(03)01368-8}
  {\path{doi:10.1016/S0168-9002(03)01368-8}}.

\bibitem{Allison:2006ve}
J.~Allison, et~al., {Geant4 developments and applications}, IEEE Trans. Nucl.
  Sci. 53 (2006) 270.
\newblock \href {https://doi.org/10.1109/TNS.2006.869826}
  {\path{doi:10.1109/TNS.2006.869826}}.

\bibitem{Allison:2016lfl}
J.~Allison, et~al., {Recent developments in Geant4}, Nucl. Instrum. Meth. A835
  (2016) 186--225.
\newblock \href {https://doi.org/10.1016/j.nima.2016.06.125}
  {\path{doi:10.1016/j.nima.2016.06.125}}.

\bibitem{VEALE2014218}
M.~Veale, S.~Bell, D.~Duarte, A.~Schneider, P.~Seller, M.~Wilson, K.~Iniewski,
  Measurements of charge sharing in small pixel cdte detectors, Nuclear
  Instruments and Methods in Physics Research Section A: Accelerators,
  Spectrometers, Detectors and Associated Equipment 767 (2014) 218--226.
\newblock \href {https://doi.org/10.1016/j.nima.2014.08.036}
  {\path{doi:10.1016/j.nima.2014.08.036}}.

\bibitem{Xu2011EvaluationOE}
C.~Xu, M.~Danielsson, H.~Bornefalk, Evaluation of energy loss and charge
  sharing in cadmium telluride detectors for photon-counting computed
  tomography, IEEE Transactions on Nuclear Science 58~(3) (2011) 614--625.
\newblock \href {https://doi.org/10.1109/TNS.2011.2122267}
  {\path{doi:10.1109/TNS.2011.2122267}}.

\bibitem{Krzyzanowska:2018}
A.~Krzyżanowska, P.~Otfinowski, P.~Gryboś, Simulations of high count rate
  performance of hybrid pixel detectors with algorithms dealing with charge
  sharing, in: 2018 International Conference on Signals and Electronic Systems
  (ICSES), 2018, pp. 23--26.
\newblock \href {https://doi.org/10.1109/ICSES.2018.8507325}
  {\path{doi:10.1109/ICSES.2018.8507325}}.

\bibitem{DTSeamlessTile}
M.~Kauppinen, A.~Winkler, V.~L\"{a}ms\"{a}, M.~Matikkala, M.~Zoladz, P.~Grybos,
  R.~Kleczek, P.~Kmon, R.~Szczygiel, T.~Fabritius, Characterization of seamless
  {CdTe} photon counting {X}-ray detector, IEEE Transactions on Instrumentation
  and Measurement 70 (2021) 1--11.
\newblock \href {https://doi.org/10.1109/TIM.2021.3070615}
  {\path{doi:10.1109/TIM.2021.3070615}}.

\bibitem{4437135}
K.~{Iniewski}, H.~{Chen}, G.~{Bindley}, I.~{Kuvvetli}, C.~B. {Jorgensen},
  Modeling charge-sharing effects in pixellated czt detectors, in: 2007 IEEE
  Nuclear Science Symposium Conference Record, Vol.~6, 2007, pp. 4608--4611.
\newblock \href {https://doi.org/10.1109/NSSMIC.2007.4437135}
  {\path{doi:10.1109/NSSMIC.2007.4437135}}.

\bibitem{KIM2011233}
J.~C. Kim, S.~E. Anderson, W.~Kaye, F.~Zhang, Y.~Zhu, S.~J. Kaye, Z.~He, Charge
  sharing in common-grid pixelated {CdZnTe} detectors, Nuclear Instruments and
  Methods in Physics Research Section A: Accelerators, Spectrometers, Detectors
  and Associated Equipment 654~(1) (2011) 233 -- 243.
\newblock \href {https://doi.org/10.1016/j.nima.2011.06.038}
  {\path{doi:10.1016/j.nima.2011.06.038}}.

\bibitem{DAgostini1995}
G.~D'Agostini, A multidimensional unfolding method based on bayes' theorem,
  Nuclear Instruments and Methods in Physics Research Section A: Accelerators,
  Spectrometers, Detectors and Associated Equipment 362~(2) (1995) 487--498.
\newblock \href {https://doi.org/10.1016/0168-9002(95)00274-X}
  {\path{doi:10.1016/0168-9002(95)00274-X}}.

\bibitem{DAgostini2010}
G.~D'Agostini, Improved iterative bayesian unfolding, arXiv
  preprintArXiv:1010.0632v1 [physics.data-an] (2010).
\newblock \href {https://doi.org/10.48550/arXiv.1010.0632}
  {\path{doi:10.48550/arXiv.1010.0632}}.

\bibitem{SpekPy}
G.~Poludniowski, A.~Omar, R.~Bujila, P.~Andreo, Technical note: Spekpy v2.0—a
  software toolkit for modeling x-ray tube spectra, Medical Physics 48~(7)
  (2021) 3630--3637.
\newblock \href {https://doi.org/10.1002/mp.14945}
  {\path{doi:10.1002/mp.14945}}.

\end{thebibliography}

\end{document}